# Broadband long-range thermal imaging via meta-correctors

Cameron Vo,[1,†] Owen Anderson,[1,†] Anna Wirth-Singh,[2] Rose Johnson,[3] Arka Majumdar,[2,3] And Zachary Coppens[1,*]

[1]*CFD Research Corporation, Huntsville, AL, 35806, USA*
[2]*Physics Department, University of Washington, Seattle, WA 98033, USA*
[3]*Electrical and Computer Engineering, University of Washington, Seattle, WA 98033, USA*
[†]*These authors contributed equally.*
[*]*Corresponding author: zachary.coppens@cfd-research.com*

**Abstract**

Long-range imaging in the thermal infrared band is critical for applications such as environmental monitoring, industrial inspections, and surveillance. To achieve high quality imaging, these systems typically require large apertures and many elements with complex shapes to correct aberrations, adding significant weight and cost. Large-area metasurface optics offer a promising solution for weight reduction; however, their substantial chromatic aberrations limit their effectiveness in the long-wave infrared (LWIR) band where broadband imaging is typically desired. In this work, we introduce a hybrid system comprising four refractive lenses and two all-silicon metasurface correctors (meta-correctors) to achieve high-quality, broadband thermal imaging at long range. Compared to a refractive-only assembly, our system demonstrates a three-fold contrast enhancement at the detector's half-Nyquist frequency. Testing outside the laboratory reveals noticeably sharper images, with human features clearly recognizable at distances of 250 meters. The assembly utilizes off-the-shelf refractive elements and avoids the use of germanium, which poses a supply chain risk. Our findings highlight the potential of hybrid meta-corrector systems to enable long-range, lightweight, and cost-effective LWIR imaging solutions.

## 1. Introduction

Long-wavelength infrared (LWIR) imaging plays a critical role in various applications ranging from environmental monitoring and industrial inspections to defense and surveillance [1–4]. Many of these applications benefit from long-range imaging, which typically requires optical systems with large apertures and multiple elements with complex surface profiles to correct aberrations. These elements add substantial weight and cost to the system, becoming the major cost driver when paired with cheap, uncooled detectors. Moreover, germanium, the most advantageous and widely used material in LWIR imaging systems, has recently become a supply chain risk [5,6]. This has further escalated costs and challenged the continued manufacturing and affordability of long-range LWIR imaging systems.

All-silicon metasurfaces offer a cost-effective approach for realizing LWIR imaging systems. Metasurfaces are subwavelength diffractive optics composed of periodic two-dimensional arrays of spatially varying subwavelength scatterers that manipulate the phase, amplitude, and polarization of transmitted wavefronts [7–10]. These flat optical elements can drastically reduce weight and thickness compared to conventional refractive lenses and can be fabricated using single-stage, high-throughput lithography and etching. This technology has gained substantial traction in recent years exploring their use in short-range, wide field of view LWIR imaging [11–15]. However, a major hindrance of metasurface imaging systems has been the strong chromatic aberration that reduces the broadband image quality. While metasurfaces can achieve some degree of achromatic imaging via dispersion engineering [16–18], the working bandwidth becomes increasingly narrow as the aperture and numerical aperture (NA) of the metasurface increases [19]. Both large aperture and high NA are desirable for LWIR imaging, which motivated the study of broadband, low-f/# metasurfaces with computational post processing to recover high-quality images [20]. Though promising, this method increases the computational load, which may be undesirable for applications needing low power and real time imaging.

Rather than relying purely on metasurfaces to build an optical system, metasurface aberration correctors (metacorrectors) can be included with refractive elements in a hybrid configuration to improve performance. Metacorrectors are capable of correcting the chromatic and monochromatic aberrations of refractive lenses, achieving improved broadband imaging in various wavebands [21–24]. These correctors can be especially advantageous in the LWIR where systems that exclude germanium are left with few refractive material options to correct aberrations. Additionally, the metacorrector approach reduces the number of refractive lenses with exotic surface profiles and lowers the size, weight, and cost of the designed systems. A recent study has shown a LWIR hybrid imaging system that combined a single refractive element with a single metacorrector [25]. While promising, the study featured a relatively wide field of view and small aperture, limiting the imaging capability to shorter ranges.

Here, we report a broadband hybrid system for long-range LWIR imaging. We avoided the use of germanium and complex refractive elements by employing four commercial-off-the-shelf ZnSe refractive lenses and two all-silicon metacorrectors. To the best of our knowledge, such six-element hybrid meta-refractive system has not been reported before. The elements were aligned and packaged within a custom 3D-printed housing. We measured the modulation transfer function (MTF) of the hybrid imaging system using the slanted edge method and showed a three-fold contrast enhancement at the half-Nyquist frequency compared to an optimized refractive-only system. The assembly was then used to capture images outside the lab, with the hybrid system delivering noticeably sharper results compared to the refractive-only system. Its long-range imaging capabilities were tested up to 250 meters, where human features remained clearly recognizable. Our results highlight the potential of this concept to enable long-range LWIR imaging using lightweight, low-cost assemblies that are free from supply chain risk.

## 2. Hybrid System Design

We designed a hybrid system using four refractive lenses and large polarization-insensitive metasurfaces. The metasurfaces correct both chromatic and monochromatic aberrations to improve the overall image quality. We used ZnSe refractive lenses with a broadband anti-reflection (AR) coating covering the LWIR band of 8-12µm. Designing optical systems, especially high NA systems, with a single material is challenging as chromatic aberration becomes increasingly pronounced and image quality suffers, as conceptualized in Figure 1(a). The hybrid system in Figure 1(b) overcomes this challenge by using a first metasurface to correct spherical aberration and longitudinal chromatic aberration and a second metasurface to correct coma and lateral chromatic aberration. Our approach provides image sharpening with virtually no increase in weight, an advantage that becomes increasingly important for longer range telescopic imaging systems where weight is a critical factor.

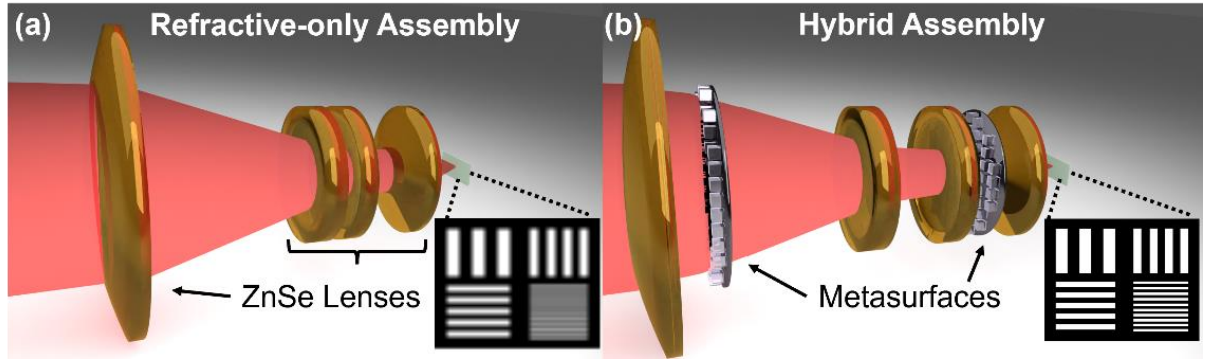

Fig. 1. Working principle for the hybrid long-range imaging system with metasurface aberration correctors. Schematics showing optimized layouts for the (a) reference refractive-only assembly and (b) hybrid system. The insets show conceptualized image sharpening and blurring for the two assemblies.

The lens assembly was designed in Ansys Zemax OpticStudio 2023 R2 to possess a narrow field of view (FOV) when paired with a FLIR Boson+ 640 camera. The Boson+ is a 640x512 format uncooled microbolometer with a 12µm pixel pitch. The system was designed for broadband performance in the 8-12µm LWIR band and was simulated and optimized for 21 equally-weighted wavelengths at six equally-weighted incident angles. The refractive elements were optimized for different surface curvatures and were later replaced with commercial off-the-shelf (COTS) lenses. The metacorrectors were modeled as Binary 2 surfaces, which add phase to the rays according to the following polynomial expansion:

$$\Phi = m \sum_{i=1}^{N} A_i \rho^{2i} \qquad (1)$$

where $m$ is the diffraction order, $\rho$ is the unitless normalized radial coordinate, $N$ is the number of polynomial coefficients in the series, and $A_i$ are the polynomial coefficients in units of radians. Eight different coefficients were used to create each metasurface phase profile.

Figure 2(a) shows the calculated broadband MTF of the hybrid system. Throughout the entire field of view, the MTF value exceeds 42% at the half-Nyquist frequency (20.8 cycles/mm) and 16% at the Nyquist frequency (41.6 cycles/mm) dictated by the pixel pitch of Boson+. The strong performance is attributed to the optimized placement of metacorrectors within the system, which is unique compared to previous meta-optic infrared hybrid systems that place metasurfaces on either side of the refractive assembly [22,25]. We note that the system was designed to have distortion in place of other optical aberrations because it can be corrected more easily via post processing. The distortion at the edge of the image was estimated to be 9.9%.

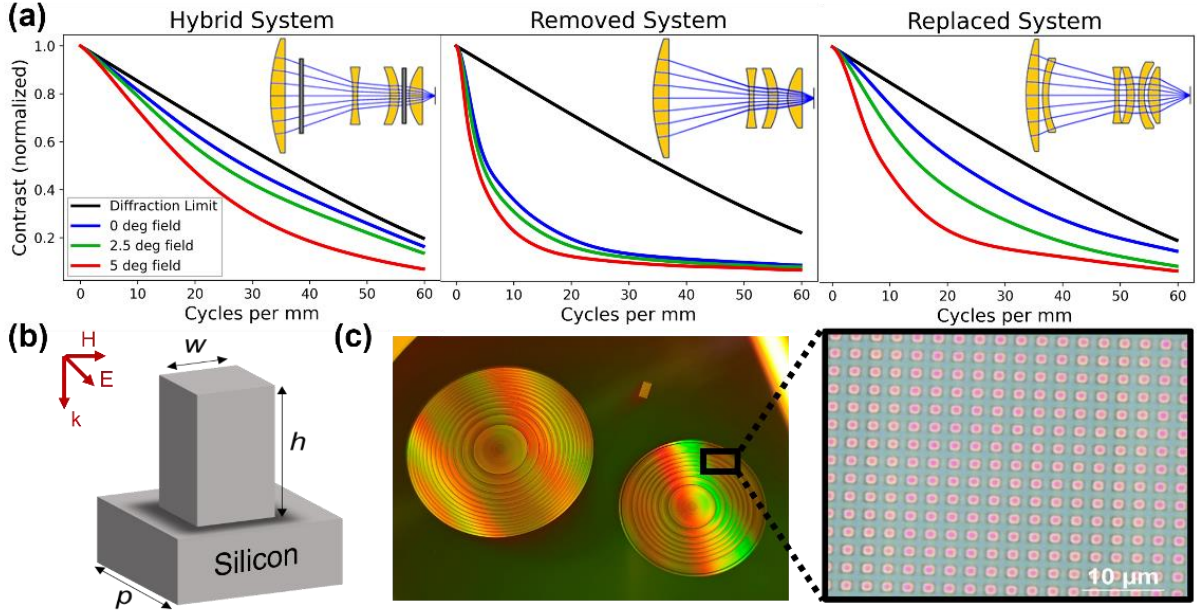

Fig. 2. Simulated MTF performance and fabrication of the metasurfaces. (a) MTF performance and ray tracing schematic for the proposed meta-optic hybrid and reference systems where the metasurfaces are removed or replaced with custom aspheric lenses. (b) Schematic of the all-silicon pillar unit cells used in the fabrication of the metasurfaces. $w$, $h$, and $p$ are the width, height, and periodicity of the pillar unit cell respectively. (c) Images of the fabricated metasurfaces taken using a standard camera and an optical microscope.

To demonstrate the impact of the metasurfaces on system performance, we created two reference systems for comparison. One system removes both metasurfaces, keeping only the COTS refractive elements, while the other system replaces both metasurfaces with custom ZnS aspheric lenses, matching the number of elements in the meta-optic hybrid. Axial position - and in the case of the custom surfaces - thicknesses, curvatures, and aspheric coefficients were optimized to maximize performance. The lens prescription data of the hybrid system and two reference systems are included in Section S1. As determined from the MTF curves shown in Figure 2(a), the hybrid meta-optic system is clearly the best performer over the whole field of view. For the replacement system using the ZnS aspheric lenses, imaging performance is mainly limited by the chromatic aberration of the ZnSe lenses. The metasurfaces have arbitrary aspheric phase profiles combined with strong negative dispersion and thus are effective for correcting the monochromatic and chromatic aberrations introduced by the ZnSe lenses. This highlights some of the challenges associated with designing germanium-free LWIR assemblies and shows the potential of metacorrectors.

The metasurfaces were designed using an all-silicon platform with square-shaped pillar meta-atoms to realize the optimized phase profiles generated from OpticStudio, which are shown in Figure S1. The phase gradient for both metasurfaces was relatively small, so the local phase approximation was used when designing the metasurface unit cell [26]. We performed a parameter sweep for different widths, periodicities, and heights using rigorous coupled-wave analysis (RCWA) and created a meta-atom library such that full 0 to $2\pi$ phase coverage was achieved for the central wavelength of 10μm. The library contained a set of 6.6μm tall pillars with 2.5μm periodicity. A schematic of the unit cell is shown in Figure 2(b), and the RCWA results are included in Section S4. It is important to note that we use a smaller period than previous LWIR metasurface lenses [25] allowing the metasurface to meet the critical sampling criteria, which eliminates resonances and lattice scattering that degrades broadband performance [27,28].

## 3. Results

The designed hybrid system was assembled and aligned using a 3D-printed lens holder as shown in Figure 3(a). Details of the holder are included in Section S5. The reference system containing only the COTS refractive lenses was also aligned using a similar lens holder. To minimize the effect of stray light from outside of the system, a shroud was attached around the holder. Additionally, an adjustable iris was mounted in front of the system and was set to a 34 mm aperture size, matching the design from OpticStudio. The assembly was mounted on a rotating stage to allow for MTF measurement across the full field of view.

We measured the MTF for the hybrid and refractive-only system using a calibrated blackbody source being shadowed by a slanted edge target. Before capture, a non-uniformity correction (NUC) and a flat-field correction (FFC) were performed to correct detector pixel non-uniformity and thermal drift common with microbolometer focal plane arrays [2]. The images of the target captured by the hybrid and refractive-only systems are shown in Figure 3(b) and 3(c), respectively. The hybrid system provides a sharper image, which was expected based on the higher MTF in the design simulations. To calculate the MTF of the system, the slanted edge images were super sampled by projecting pixel values onto a perpendicular line and an edge spread function was calculated [29]. The derivative of the edge spread function was calculated resulting in a 1-dimensional approximation of the MTF, called the line-spread function. Results of the MTF measurements are displayed in Figure 3(d).

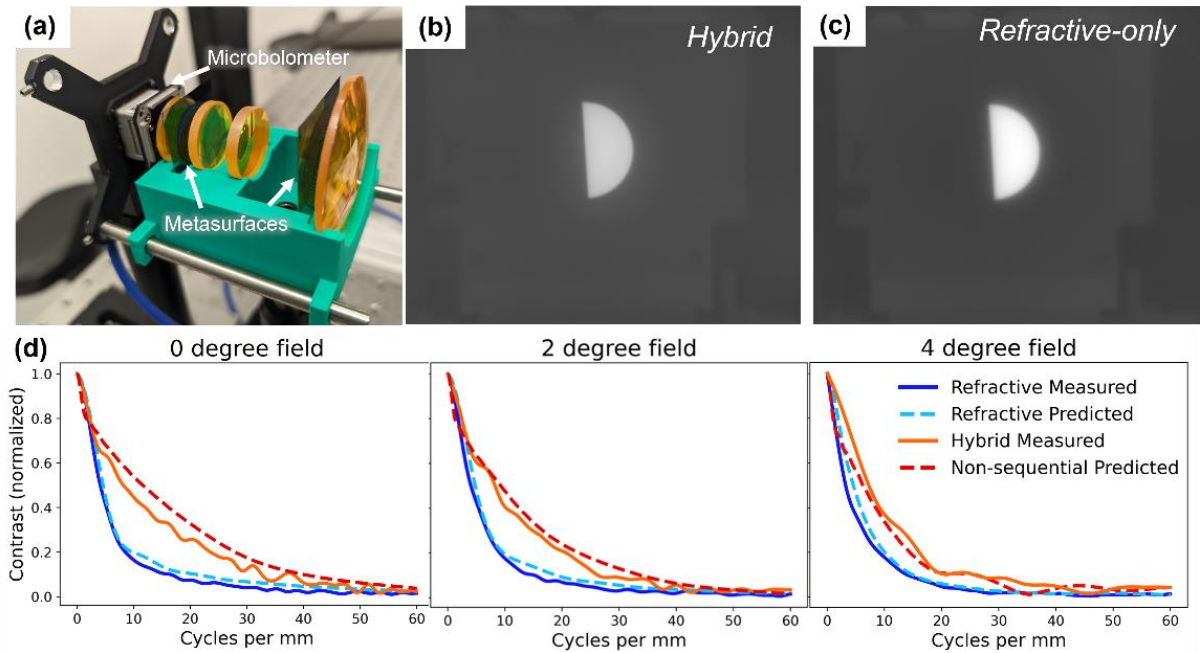

Fig. 3. Experimental setup and characterization of the hybrid and refractive-only systems. (a) 3D printed lens holder used to align the lenses and detector of the meta-optic hybrid. The adjustable iris, securing lid, and shroud used in the measurements are not shown in the image. Images of a slanted edge target are captured using (b) the meta-optic hybrid and (c) the refractive-only assembly. The blackbody temperature was set to 100°C. (d) Measured and predicted MTFs

at different field angles for the meta-optic hybrid and the refractive-only system. The predicted MTF or the hybrid is simulated via OpticStudio's non-sequential mode.

The hybrid system outperforms the refractive assembly across the field of view, providing approximately three-fold contrast enhancement at the half-Nyquist frequency of 20.8 cycle/mm. The slight degradation near the edge of the field is a result of the slanted edge target being in front of the hyperfocal distance. Though we measure substantial enhancement in image quality, our simulations in Figure 2(a) show the possibility of obtaining even more contrast improvement. The discrepancy could be due to a number of factors including errors in alignment, non-idealities in the fabricated metasurface, and diffraction efficiency losses from the meta-atom. To understand how the meta-atom diffraction affects MTF, we developed a simulation method that couples meta-atom scattering parameters with the ray tracing model. Our method employed OpticStudio's non-sequential ray tracing calculations which permit rays to travel through optical components in any diffraction order and enables the splitting, scattering, and reflecting of rays. Scattering parameters for representative meta-atom gratings were calculated using RCWA and input to the model to calculate the geometric MTF. The geometric MTF is an approximation that is recommended when scattering increases background illumination on the focal plane, as is the case with our hybrid system. More details of the investigation are included in Section S6.

Figure 3(d) shows the simulated geometric MTF accounting for the meta-atom diffraction efficiency. The results closely match the measured hybrid system MTF throughout the full FOV, including the sharp decline in MTF at low frequencies. This decline is common for hybrid systems and is caused by poor focusing of higher and lower diffraction orders scattered by the diffractive optic that contribute to a uniform background illumination [30]. With confidence in our new modelling framework, we investigated meta-atom design alterations that could improve diffraction efficiency and boost MTF performance (see Section S6). Improved MTF results over the full field were obtained when switching from a pillar geometry to a hole, as the pillar possesses a larger amount of undesired phase dispersion that lowers broadband diffraction efficiency. Additionally, the hole layout allows for the inclusion of an anti-reflection coating [13], which improves average transmission efficiency from 80.4% to 93.1%. We plan to investigate hole meta-correctors in future work.

After characterizing the imaging system in the lab, the hybrid system was taken outside to demonstrate performance in a more realistic and less-controlled environment. The images shown in Figure 4 were taken near the CFD Research building when outside temperatures were approximately 29º C. The scenes were chosen to highlight the unique capabilities of our system. We show in the left panel images of humans on a roof taken at a range of 250m. It is clear that the hybrid system allows for recognition of human features, while the refractive system shows soft edges and indistinguishable features. In the middle panel, we again see that the hybrid system outperforms by providing high spatial frequency information of the building and identification of a small drone out to 140 m. While the drone shows low contrast, it is discernable with both the human eye and software. In contrast, no moving object was perceptible when imaging with the refractive-only system. For a final demonstration, the right panel shows sharpening of the lettering on the building with the hybrid system. We note that no post processing other than distortion correction was performed on the images. The results of this outdoor testing validate our design concept and demonstrate that a meta-optic hybrid system can provide high-quality, long-range imaging in a lightweight, low-cost assembly.

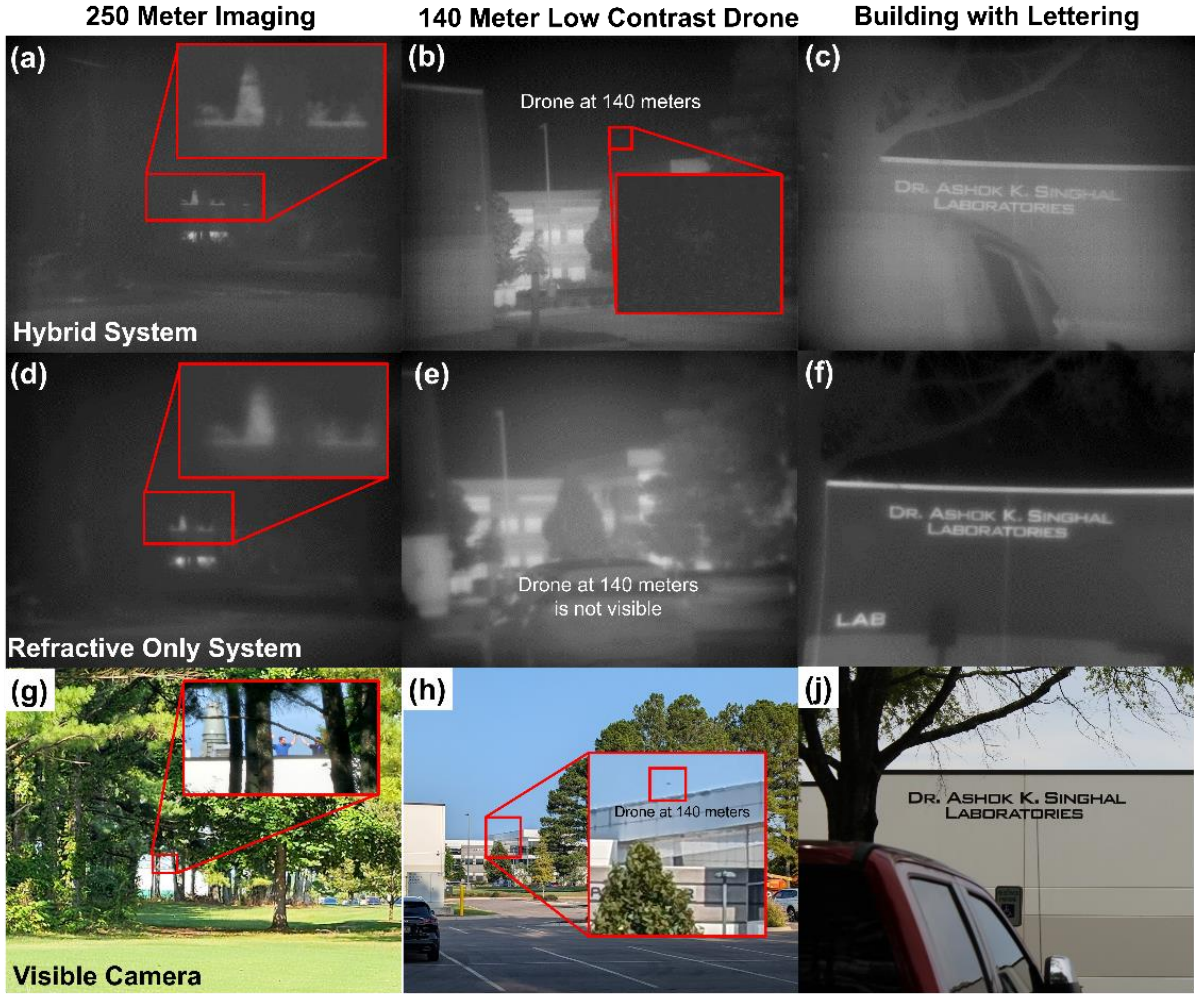

Fig. 4. Thermal and visible images captured outside of the laboratory. Images taken with the meta-optic hybrid of (a) two people on the roof of a building at 250 meters, (b) a drone at 140 meters, and (c) the side of a building with lettering. Comparison images are provided using (d)—(f) the refractive-only assembly and (g)—(j) a visible camera.

## 4. Discussion

In this work, we designed, fabricated, and tested a hybrid system composed of four ZnSe COTS lenses and two metasurface optics for broadband LWIR imaging. Our concept provides a viable solution for realizing lightweight, long-range LWIR imaging systems without using germanium elements, which are becoming a supply chain risk. When compared to the refractive-only assembly, our meta-optic hybrid system demonstrated three-fold contrast enhancement at the half-Nyquist frequency. We also showed high-quality imaging under ambient conditions outside of the laboratory with clear recognition of humans out to 250 meters. Additional advantages of our design include the considerably smaller thermo-optic coefficient of silicon metacorrectors compared to germanium, resulting in a higher tolerance to large temperature changes [31,32]. Also, all-silicon metasurfaces can be manufactured at low cost and high rates due to the well-established silicon foundry infrastructure.

In future work, we plan to switch to an AR-coated hole meta-corrector as we found improved diffraction and transmission efficiencies with this unit cell geometry. We also plan to improve imaging performance by optimizing the system using our newly developed modeling approach that accounts for meta-atom dispersion. Recent work has shown that high fidelity imaging systems can be achieved through proper optimization of meta-refractive hybrid systems [33]. Overall, meta-optic hybrid systems hold promise to opening new avenues for developing advanced long-range LWIR imaging systems, with potential applications in areas such as remote sensing, perimeter security, and environmental observation.

**Funding.** Funding for this work was supported by the federal SBIR program. Part of this work was conducted at a National Nanotechnology Coordinated Infrastructure (NNCI) site at the University of Washington with partial support from the National Science Foundation.

**Data availability.** Data underlying the results presented in this paper are not publicly available at this time but may be obtained from the authors upon reasonable request.

# Supplementary Material

## S1. Lens Prescription Data

The prescriptions of the meta-optic hybrid and two reference systems shown in Figure 2(a) are listed in Tables S1, S2, and S3. The Binary 2 surface parameters used to define the phase profiles of the metasurfaces can be found in Table S4. $\rho$ is the normalization radius while D is the diameter of the metasurface. The phase profiles of each metasurface are shown in Figure S1.

**Table S1. Lens prescription for the meta-optic hybrid**

| Surface | Part No. | Radius (mm) | Thickness (mm) | Material | Conic | 4$^{th}$ | 6th | 8$^{th}$ | 10th |
|---|---|---|---|---|---|---|---|---|---|
| 1 | Edmund Optics #39-518 | 7.13E+01 | 6.50E+00 | ZNSE | -1.28E+00 | -1.50E-07 | -1.97E-11 | 6.33E-15 | |
| 2 | | Infinity | 6.73E+00 | AIR | | | | | |
| 3 | Metasurface 1 | Infinity | 5.00E-01 | SILICON | | | | | |
| 4 | | Infinity | 2.27E+01 | AIR | | | | | |
| 5 | Thorlabs LD7671-E3 | -7.18E+01 | 2.00E+00 | ZNSE | | | | | |
| 6 | | 7.18E+01 | 1.44E+01 | AIR | | | | | |
| 7 | Thorlabs LE7031-E3 | -3.26E+01 | 4.00E+00 | ZNSE | | | | | |
| 8 | | -2.84E+01 | 2.00E+00 | AIR | | | | | |
| 9 | Metasurface 2 | Infinity | 5.00E-01 | SILICON | | | | | |
| 10 | | Infinity | 2.00E+00 | AIR | | | | | |
| 11 | Edmund Optics #39-510 | 1.78E+01 | 6.00E+00 | ZNSE | -1.24E+00 | -5.27E-06 | -3.69E-08 | 1.27E-10 | -1.48E-13 |
| 12 | | Infinity | 5.50E+00 | AIR | | | | | |

**Table S2. Lens prescription for the reference system where the metasurfaces are removed**

| Surface | Part No. | Radius (mm) | Thickness (mm) | Material | Conic | 4$^{th}$ | 6th | 8th | 10th |
|---|---|---|---|---|---|---|---|---|---|
| 1 | Edmund Optics #39-518 | 7.13E+01 | 6.50E+00 | ZNSE | -1.28E+00 | -1.50E-07 | -1.97E-11 | 6.33E-15 | |
| 2 | | Infinity | 3.34E+01 | AIR | | | | | |
| 3 | Thorlabs LD7671-E3 | -7.18E+01 | 2.00E+00 | ZNSE | | | | | |
| 4 | | 7.18E+01 | 6.00E+00 | AIR | | | | | |
| 5 | Thorlabs LE7031-E3 | -3.26E+01 | 4.00E+00 | ZNSE | | | | | |
| 6 | | -2.84E+01 | 4.00E+00 | AIR | | | | | |
| 7 | Edmund Optics #39-510 | 1.78E+01 | 6.00E+00 | ZNSE | -1.24E+00 | -5.27E-06 | -3.69E-08 | 1.27E-10 | -1.48E-13 |
| 8 | | Infinity | 4.97E+00 | AIR | | | | | |

**Table S3. Lens prescription for the reference system where the metasurfaces are replaced**

| Surface | Part No. | Radius (mm) | Thickness (mm) | Material | Conic | 4th | 6th | 8th | 10th |
|---|---|---|---|---|---|---|---|---|---|
| 1 | Edmund Optics #39-518 | 7.13E+01 | 6.50E+00 | ZNSE | -1.28E+00 | -1.50E-07 | -1.97E-11 | 6.33E-15 | |
| 2 | | Infinity | 2.00E+00 | AIR | | | | | |
| 3 | Custom Lens 1 | 4.81E+01 | 3.00E+00 | ZNS | -1.30E+00 | -1.38E-06 | -3.02E-09 | 1.22E-11 | -2.17E-14 |
| 4 | | 3.55E+01 | 3.42E+01 | AIR | | | | | |
| 5 | Thorlabs LD7671-E3 | -7.18E+01 | 2.00E+00 | ZNSE | | | | | |
| 6 | | 7.18E+01 | 4.00E+00 | AIR | | | | | |
| 7 | Thorlabs LE7031-E3 | -3.26E+01 | 4.00E+00 | ZNSE | | | | | |
| 8 | | -2.84E+01 | 2.00E+00 | AIR | | | | | |
| 9 | Custom Lens 2 | 5.70E+01 | 3.00E+00 | ZNS | -1.30E+00 | 2.74E-05 | 7.37E-08 | -3.66E-10 | 5.54E-13 |
| 10 | | 2.23E+01 | 2.00E+00 | AIR | | | | | |
| 11 | Edmund Optics #39-510 | 1.78E+01 | 6.00E+00 | ZNSE | -1.24E+00 | -5.27E-06 | -3.69E-08 | 1.27E-10 | -1.48E-13 |
| 12 | | Infinity | 1.56E+01 | AIR | | | | | |

**Table S4. Binary 2 parameters for the metasurfaces**

| Metasurface | $\rho$ (mm) | D (mm) | A1 | A2 | A3 | A4 | A5 | A6 | A7 | A8 |
|---|---|---|---|---|---|---|---|---|---|---|
| Metasurface 1 | 18.00 | 32.00 | -88.132 | 0.505 | -338.475 | 2041.012 | -6428.354 | 11042.71 | -9859.70 | 3660.60 |
| Metasurface 2 | 12.00 | 24.00 | -43.93 | -282.968 | 2107.63 | -8747.905 | 19784.44 | -24948.515 | 16612.35 | -4583.42 |

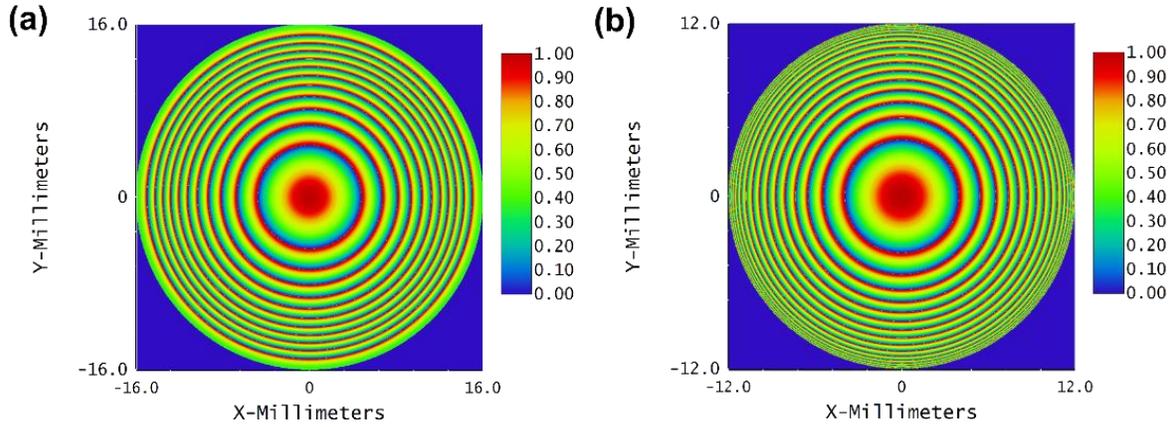

Fig. S1. Metasurface phase profiles. Phase profiles of (a) metasurface 1 and (b) metasurface 2.

## S2. Hybrid system optimization

The hybrid system was first optimized in OpticStudio using a merit function that minimized root mean square (RMS) spot size across the full field of view. The damped least squares method was used for the initial, course optimization. To further finetune the optimization, we switched to a different merit function that used OpticStudio's Contrast Optimization method, which optimizes the MTF at a specified spatial frequency. We chose to optimize the system at 20.8 cycles/mm which is the half-Nyquist frequency of the FLIR Boson+ 640 detector.

## S3. Metasurface fabrication

The metacorrectors were fabricated on a 300μm thick double side polished silicon wafer. To promote photoresist adhesion, we first dehydrated and deposited hexamethyldisilazane (HMDS) in a YES-HMDS oven. A positive photoresist (AZ 1505) was then spin coated at 4000 rpm, followed by a bake at 100°C for 90 seconds. The resist was exposed using direct write lithography (Heidelberg DWL - 66+), baked post-exposure at 100°C for 60 seconds, and then developed in AZ-726 MiF developer. Using the photoresist as a mask, the silicon was etched to the 6.6 μm depth using deep reactive ion etching (SPTS Rapier). After etching, oxygen plasma (YES CV200 RFS) was used to strip residual photoresist. To maintain fabrication consistency between the two metasurfaces, both were fabricated on a single wafer that was diced (Disco - DAD321) afterwards to separate the metasurfaces. A thick photoresist (AZ 10XT) was used to protect the structures during dicing and was later removed with acetone and IPA.

## S4. Meta-atom RCWA simulation data

The RCWA simulation results for the pillar unit cell library used to fabricate the metasurfaces are shown in Figure S2. The normalized phase response and transmission through the unit cell at the wavelengths of 8μm, 10μm, and 12μm are shown. The fabricated metasurfaces were discretized using 13 different pillar widths. Due to the high reflectivity of silicon metasurfaces and relatively large phase dispersion of pillars, we have also investigated using hole meta-atoms with a single layer ZnS AR coating with results shown in Figure S3. We designed a unit cell library using holes with a fixed depth of 7.5μm, 2.5μm periodicity, and an additional 1.19 μm thick AR coating. The AR coating increases average transmission from 80.4% to 93.1% through the unit cell. Unlike the pillars, we selected the widths of the holes such that full 0 to 2π phase coverage was achieved for 9μm light. This is because we have observed that shorter wavelengths lose diffraction efficiency at an increased rate compared to wavelengths longer than the design wavelength.

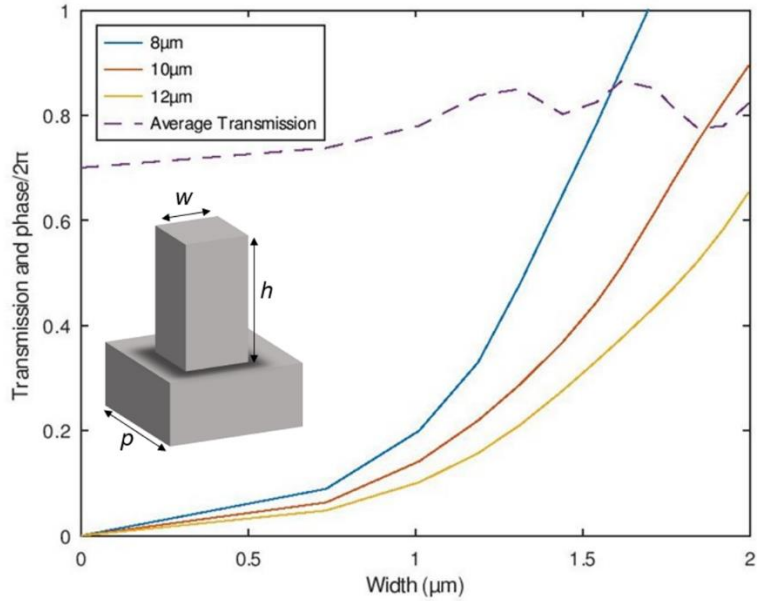

Fig. S2. Phase and transmission data for the pillar unit cell library. The phase data is shown for three different wavelengths. The transmission data shown is the average transmission across the three wavelengths.

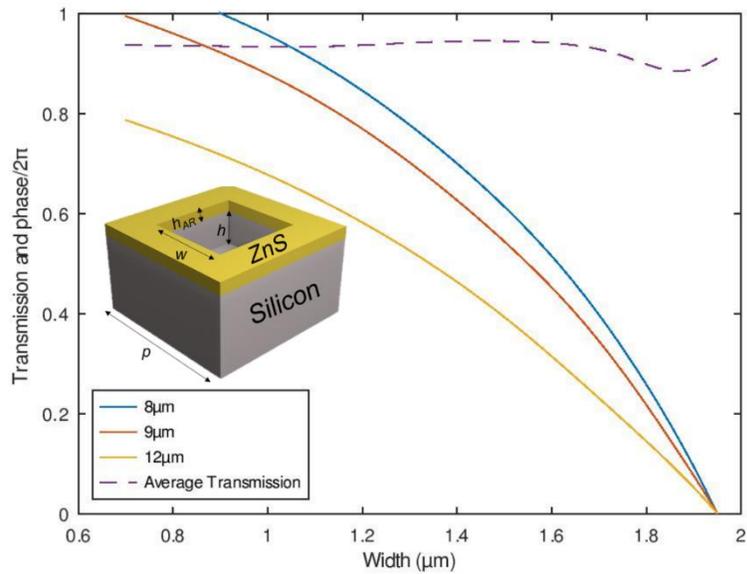

Fig. S3. Phase and transmission data for the AR-coated hole unit cell library. $h$, $w$, and $p$ are the depth, width, and periodicity of the hole structures and $h_{AR}$ is the thickness of the AR coating.

To compare the performance of holes and pillars, we designed two gratings consisting of pillar and hole unit cells such that a first order diffraction angle of 0.2° was achieved at their respective design wavelengths. The pillar-based grating used the 13 discretized widths from the fabricated metasurfaces, while the hole-based grating used 16 discretized widths that were selected from the designed unit cell library shown in Figure S3. Since the phase gradients of the metacorrectors were small and slowly varied throughout the aperture, we assumed that the diffraction efficiency throughout the metasurfaces was independent of radial position and would only vary with wavelength.

## S5. Characterization and imaging setup

The focal plane array used was the FLIR Boson+ 640, which has a detector size of 512 by 640 pixels and a pixel pitch of 12 µm. The detector was placed in a custom-machined mount that connected to the lens holders with rods. The rods aligned the optical system with the detector while allowing for the lens system to travel towards and away from the detector for focusing. The lens holder assembly consisted of a top and a bottom section for both the refractive and hybrid optical systems. The assemblies were made in SolidWorks software and printed in acrylonitrile butadiene styrene (ABS) with the Fusion3 Edge 3D printer. Indentations of the lenses were made by subtracting 3D models of the lenses from a solid shape. The metasurfaces were held in place with small bricks extending from the top, bottom, and one side of the model, which can be seen in Figure S4(a). The two halves of the assembly were clamped or taped together to hold the system in place.

The lens holder was designed to not apply clamping pressure to the metasurfaces, because they are brittle and break easily. There is a screw hole in the bottom of the lower piece so that the lenses can be affixed to a standard ¼-20 optical post. The post is attached to a fine-motion translation stage for focusing. The horizontal post holes in the lower piece are designed to be loose to allow for the lens assembly to smoothly slide in and out from the detector while maintaining alignment. When the lens assemblies were used for testing the system was held tightly together with tape that also acted as a stray light baffle, which is shown in Figure S4(c).

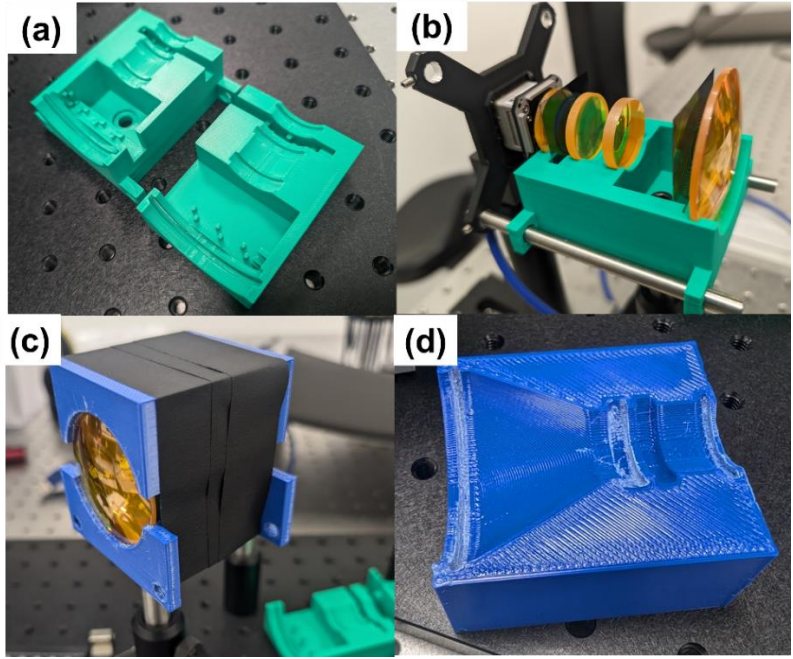

Fig. S4. 3D printed mounts for the hybrid and refractive-only systems. (a) Hybrid system holder bottom (left) and top (right). (b) Hybrid lens holder system without top, connected to the Boson 3+ with optical posts. (c) Refractive lens assembly with black painter's tape to hold the system for testing. (d) Top of the refractive holder.

The lens assembly was attached to a translation stage for fine adjustment focus. Focusing was performed by visual inspection as well as live MTF calculations of the slanted edge target. MTF measurements were performed for the hybrid and refractive-only system by directly imaging Santa Barbara Infrared 4" Differential Blackbody being shadowed by a slanted edge target. The blackbody and edge target were positioned 4.7 m away from the imaging system. This is closer than the hyperfocal distance and, according to simulations, results in reduced off-axis imaging performance. The hybrid and refractive system were mounted to a breadboard so that they could be taken outside of the laboratory for outdoor imaging.

## S6. Diffraction efficiency loss in OpticStudio

Using the simulated grating data, we analyzed the effects of diffraction efficiency losses on the hybrid system's performance in OpticStudio's non-sequential mode. Non-sequential ray tracing permits rays to travel through optical components in any order and enables the splitting, scattering, and reflecting of rays. For diffractive surfaces, the relative amount of energy that is split into each transmissive and reflective order can also be specified. We evaluated the hybrid system at 8µm, 9µm, 10µm, 11µm, and 12µm, creating a configuration for each wavelength. In each configuration, the grating transmission and reflection data for each wavelength was input into the diffraction properties of both metasurfaces, ensuring that diffraction efficiency losses were accounted for.

Rays were traced through each configuration and incoherently summed on the detector for each field angle. The geometric MTF was calculated and then multiplied by the diffraction limited MTF to approximate diffraction effects at the focal plane. The MTF results for the ideal hybrid system and systems with hole and pillar unit cells are shown in Figure S5. The ideal system possesses 100% diffraction efficiency into the 1$^{st}$ order and represents a baseline when using the geometric

MTF approximation. The MTF performance for the hole system is noticeably improved at lower spatial frequencies, which is expected considering the decreased phase dispersion of the unit cell. Additionally, since the holes are AR coated, reflection losses and stray light are reduced. Overall, switching to AR coated holes from pillars improves image quality, nearly matching the ideal performance.

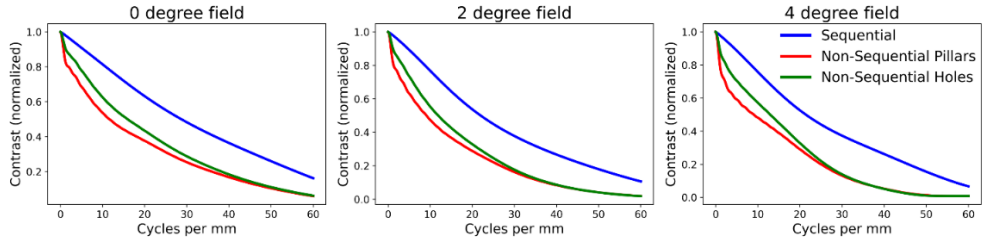

Fig. S5. MTF of the hybrid system when including diffraction losses. The MTF is calculated in nonsequential mode when using hole and pillar unit cells. The MTF calculated from sequential mode is included for reference.